\documentclass[prb,twocolumn,english,superscriptaddress]{revtex4-1}

\usepackage[T1]{fontenc}
\usepackage{soul}

\usepackage{amsmath}
\usepackage{mathtools}
\usepackage{amsfonts}

\usepackage{color}
\usepackage{xcolor}
\usepackage[colorlinks,
            citecolor=blue,
            linkcolor=blue,
            urlcolor=blue]{hyperref}

\usepackage{graphicx,placeins}
\usepackage[all]{hypcap} 

\begin{document}

\title{Evidence of two-spinon bound states in the magnetic spectrum of Ba$_3$CoSb$_2$O$_9$}

\author{E. A. Ghioldi}
\affiliation{Department of Physics and Astronomy, University of Tennessee, Knoxville, Tennessee 37996, USA}
\affiliation{Instituto de Física Rosario (CONICET) and Universidad Nacional de Rosario, Boulevard 27 de Febrero 210 bis, (2000) Rosario, Argentina}

\author{Shang-Shun~Zhang}
\affiliation{Department of Physics and Astronomy, University of Tennessee, Knoxville, Tennessee 37996, USA}
\affiliation{School of Physics and Astronomy and William I. Fine Theoretical Physics Institute, University of Minnesota, Minneapolis, Minnesota 55455, USA}
	
\author{Yoshitomo Kamiya}
\affiliation{School of Physics and Astronomy, Shanghai Jiao Tong University, Shanghai 200240, China}

\author{L. O. Manuel}
\affiliation{Instituto de Física Rosario (CONICET) and Universidad Nacional de Rosario, Boulevard 27 de Febrero 210 bis, (2000) Rosario, Argentina}

\author{A. E. Trumper}
\affiliation{Instituto de Física Rosario (CONICET) and Universidad Nacional de Rosario, Boulevard 27 de Febrero 210 bis, (2000) Rosario, Argentina}

\author{C. D. Batista} 
\email{cbatist2@utk.edu}
\thanks{Corresponding author}
\affiliation{Department of Physics and Astronomy, University of Tennessee, Knoxville, Tennessee 37996, USA}
\affiliation{Shull Wollan Center - A Joint Institute for Neutron Sciences, Oak Ridge National Laboratory, Oak Ridge, Tennessee 37831, USA}

\begin{abstract} 
    
    Recent inelastic neutron scattering (INS)  experiments of the triangular antiferromagnet Ba$_3$CoSb$_2$O$_9$ 
    revealed strong deviations from semiclassical theories. We demonstrate that key features of the INS data are 
    well reproduced by a parton Schwinger boson theory beyond the saddle-point approximation. The measured magnon 
    dispersion is well reproduced by the dispersion of two-spinon bound states 
    (poles of the emergent gauge fields propagator), while the low-energy  continuum scattering
    is reproduced by a quasifree two-spinon continuum, suggesting that a free spinon gas is a good initial 
    framework to study magnetically ordered states near a quantum melting point.
    
\end{abstract}

\maketitle


\section {Introduction} 

Identifying new states of matter is a central theme of condensed matter physics. Although theorists have predicted an abundance of such states, it is often difficult to find experimental realizations. This is particularly a challenge for quantum spin liquids (QSLs), where the lack of smoking-gun  signatures  is forcing the community to develop more comprehensive approaches~\cite{Knolle2019_review,broholm2019quantum,Savary_2016review}. The singular interest in the fractionalized quasi-particles of these highly entangled states of matter resides in their potential application to quantum information~\cite{Kitaev03,broholm2019quantum,tokura17}. However, it has been frustratingly difficult to detect these quasiparticles in real materials.

Since Anderson's proposal of the resonating valence bond state~\cite{Anderson1973}, 
the triangular geometry has long been studied as a platform for finding QSLs.
Although the ground state of the simplest spin-$1/2$ model with nearest-neighbor (NN) antiferromagnetic Heisenberg interactions $J_1$ exhibits a $120^{\circ}$ long-range magnetic order, geometric frustration makes this order weak~\cite{Capriotti_1999,White07}. Indeed, a next-nearest-neighbor exchange coupling $J_2$ as small as $\approx 0.06 J_1$ is enough to continuously melt the magnetic order into a QSL phase~\cite{Hu15,Iqbal16,Saadatmand16,Wietek17,Gong17,Hu19,Zhu15,Zhu18}. Determining the nature of the QSL  is an ongoing theoretical challenge, with proposals ranging from gapped $\mathbb{Z}_2$ and gapless $U(1)$ Dirac to chiral~\cite{Zhu15,Hu15,Iqbal16,Saadatmand16,Wietek17,Gong17,Hu19}. To discern among QSL candidates and the corresponding low-energy parton theories, it is imperative to make contact with experiments. 
Since most of the known realizations of the triangular lattice Heisenberg  antiferromagnet (TLHA) lie on the ordered side of the quantum critical point (QCP) at $J_2/J_1\approx 0.06$~\cite{Manuel99,Mishmash13,Kaneko14,Li15,Zhu15,Zhu18}, reproducing their measured excitation spectrum 
is the most stringent test for alternative parton theories.  

The idea of describing two-dimensional (2D) frustrated antiferromagnets by means of fractional excitations (spinons) coupled to emergent gauge fields has been around for many years~\cite{Arovas88,Sachdev1991,Read1991,Chubukov94,Chubukov1996}. The Schwinger boson theory (SBT) is one of the  first parton formulations that was introduced to describe ordered and disordered phases on an equal footing
~\cite{Auerbach94,Read1991}. However, a qualitatively correct, beyond the saddle-point (SP) level, computation of the dynamical structure factor of {\it magnetically ordered} phases has been achieved only recently~\cite{Ghioldi18,Zhang19,Zhang21}, enabling comparisons with 
inelastic neutron scattering (INS) measurements.
 
Ba$_3$CoSb$_2$O$_9$ is one of the best known realizations of a spin-$1/2$ TLHA \cite{Macdougal20,Zhou_2012,Ito2017,Ma16}. INS studies of this material~\cite{Ma16,Ito2017,Macdougal20} reveal an unusual  three-stage energy structure of the magnetic spectral weight [see Fig. \ref{flo:SchwingerBosons2}(a)]. The lowest-energy 
stage is composed of dispersive branches of single-magnon excitations. The second and third stages correspond to dispersive continua that extend up to energies  six times larger than the single-magnon bandwidth~\cite{Ito2017}. 
These observations are quantitatively and qualitatively inconsistent with nonlinear spin wave theory (NLSWT)~\cite{Ma16,Kamiya18}, suggesting  that magnons could be better described as two-spinon bound states
(spinons are the fractionalized quasiparticles 
of the neighboring QSL state).  Here we investigate this hypothesis by comparing the INS data of 
Ba$_3$CoSb$_2$O$_9$~\cite{Ito2017,Macdougal20} against the SBT described in Refs.~\cite{Ghioldi18,Zhang19,Zhang21}.

These comparisons 
 demonstrate that  a low-order SBT provides 
 an adequate starting point to reproduce the measured spectrum of low-energy excitations, including the magnon dispersion  (first stage) reported in Ref.~\cite{Macdougal20} and the dispersion of the broad low-energy peak that appears in the continuum (second stage)~\cite{Ito2017,Macdougal20}. Importantly, these results  
 shed light on the nature of the proximate QCP 
 and of the quantum spin liquid phase that is expected for $J_2/J_1 \gtrsim 0.06$~\cite{Scheie21}.

\section{Material and Model} 

Ba$_3$CoSb$_2$O$_9$  comprises vertically stacked triangular layers of effective spin-1/2 moments arising from the $\mathcal{J} = 1/2$ Kramers doublet of Co$^{2+}$ in  a trigonally-distorted octahedral ligand field. Excited multiplets are separated by a gap of 200-300 K due to spin-orbit coupling, which is much larger than the N\'eel temperature $T_\text{N} = 3.8$ K. Below $T_\text{N}$,
the material develops conventional 120$^\circ$ ordering with wavevector $\mathbf{Q} = (1/3,1/3,1)$~\cite{Doi2004}. 
The theoretical modeling of different experimental results~\cite{Susuki2013,Koutroulakis2015,Ma16,Kamiya18} indicates that the magnetic properties of Ba$_3$CoSb$_2$O$_9$ are well described by the \textit{XXZ}  model:
\begin{equation}
  {\mathcal H} = 
   \sum_{\langle{i,j}\rangle} J_{ij} \left( S^x_i S^x_j + S^y_i S^y_j + \Delta S^z_i S^z_j \right),
  \label{eq:model}
\end{equation}
\noindent where $\langle i, j\rangle$  restricts the sum to   NN intralayer and interlayer  bonds with exchange interactions  $J_{ij}=J$ and $J_{ij}=J_c$, respectively, and $\Delta$ accounts for a small easy plane exchange anisotropy
~\footnote{The high-symmetry structure of this material forbids  Dzyaloshinskii-Moriya interactions between  Co$^{2+}$ ions in the same $ab$ plane or relatively displaced along the $c$-axis}.
Since the set of in-plane Hamiltonian parameters reported in Refs.~\cite{Ito2017,Macdougal20,Kamiya18} coincide with each other within a relative error
$\sim 5\%$, here we adopt the values $J=1.66$ meV  and $\Delta=0.937$. As for the inter-plane exchange, we adopt $J_c=0.061J$, between the values $J_c=0.05J$ and $J_c=0.08J$ reported in Refs.~\cite{Kamiya18} and \cite{Ito2017,Macdougal20}, respectively. As expected for this effective spin model, experiments confirmed a
one-third magnetization plateau (up-up-down phase) induced by a magnetic field parallel to the easy-plane~\cite{Chubukov91,Shirata2012,Susuki2013,Koutroulakis2015,Quirion2015,Sera2016}. While the dynamical spin structure factor of the up-up-down phase is well described by NLSWT~\cite{Alicea09,Kamiya18},
the observed zero-field magnon dispersions cannot be described with any known semiclassical treatment~\cite{Ma16,Ito2017}, suggesting that quantum renormalization effects in the zero field are underestimated by a perturbative $1/S$ expansion.
These strong quantum fluctuations can be attributed  to the proximity of 
the TLHA to the above-mentioned  ``quantum melting point'' that signals a \emph{continuous} $T=0$ transition  into a quantum spin liquid.

\section{Schwinger Boson Theory} 

The SBT~\cite{Arovas88,Auerbach94,Ghioldi18} starts from a parton representation of the spin operators expressed in terms of spin-$1/2$ bosons that represent the spinons of the theory:
$\hat{\boldsymbol S}_i = \frac{1}{2}  {\boldsymbol b}_{i}^{\dag}   {\boldsymbol \sigma}  {\boldsymbol b}_{i}$, 
where ${\boldsymbol b}_{i}^{\dag}=(b_{i\uparrow}^{\dag}, \ b_{i\downarrow}^{\dag} )$, and ${\boldsymbol \sigma} \equiv (\sigma^x, \ \sigma^y, \ \sigma^z)$ is the vector of Pauli matrices. The spin-$1/2$ representation of the spin operator is enforced by  the constraint $ \sum_{\sigma=\uparrow, \downarrow} b_{i \sigma}^{\dag}b_{i \sigma}^{}=1$.
The advantage of this representation is that the spin-spin interaction can be expressed as a bilinear form, $X^{\dagger}_{ij}X_{ij}$, in  bond operators $X_{ij}$ which are invariant under the spin-rotation symmetries of the Hamiltonian. Correspondingly,  the  mean-field approximation $X^{\dagger}_{ij}X_{ij} \simeq \langle X^{\dagger}_{ij} \rangle X_{ij} + X^{\dagger}_{ij} \langle X_{ij} \rangle - \langle X^{\dagger}_{ij} \rangle \langle X_{ij} \rangle $  preserves the rotational symmetry of the spin Hamiltonian. This is one of the important differences between SBT and  spin wave theory~\cite{Arovas88,Auerbach94}.

For the case of interest, the \textit{XXZ} interaction [Eq.~\eqref{eq:model}] can be expressed in terms of SU(2) spin-rotation invariant bond operators \cite{Ghioldi2015}, $ A_{ij}^{} = \frac{1}{2} (b_{i\uparrow}b_{j\downarrow} - b_{i\downarrow}b_{j\uparrow})$, $ B_{ij}^{} = \frac{1}{2} (b_{j\uparrow}^{\dag} b_{i\uparrow}^{} + b_{j\downarrow}^{\dag} b_{i\downarrow}^{})$,  and U(1) spin-rotation invariant bond operators $ C_{ij}^{} = \frac{1}{2} (b_{j\uparrow}^{\dag} b_{i\uparrow}^{} - b_{j\downarrow}^{\dag} b_{i\downarrow}^{})$  and  $ D_{ij}^{} = \frac{1}{2} (b_{i\uparrow}^{} b_{j\downarrow}^{} + b_{i\downarrow}^{} b_{j\uparrow}^{})$ required to account for the finite uniaxial anisotropy. The operator $A^{\dagger}_{ij}$ ($D^{\dagger}_{ij}$) creates a singlet (triplet) state on the bond $ij$.
The operator $B_{jk}$ moves  singlets and triplets from the bond $ij$ to the bond $ik$ preserving their character. In contrast, the operator  $C_{jk}$ promotes a singlet bond $ij$ into a triplet bond $ik$, and vice versa. Up to an irrelevant constant, the spin-spin interaction is expressed as~\cite{Scheie21} 
\begin{align}
   S^x_i S^x_j + S^y_i S^y_j + \Delta S^z_i S^z_j  = - 2\left(\frac{\Delta+1}{2}-\alpha \right) A_{ij}^{\dag} A_{ij}^{} \nonumber \\
   +  2\alpha :\! B_{ij}^{\dag} B_{ij}^{}\!: +  \frac{\Delta-1}{2} \;(:\!C_{ij}^{\dag} C_{ij}^{}\!: - D_{ij}^{\dag} D_{ij}^{}).
   \label{spinspin}
\end{align}
The continuous parameter $\alpha$  parameterizes equivalent ways of expressing the spin-spin interaction by assigning different weights to  $A_{ij}^{\dag} A_{ij}^{}$ and $:\! B_{ij}^{\dag} B_{ij}^{}\!:$. This parametrization leads to a family of possible mean-field solutions in the canonical formalism or Hubbard-Stratonovich transformations in the path-integral formulation. Similarly to the case of KYbSe$_2$~\cite{Scheie21}, the optimal value of $\alpha$ is obtained by fitting the INS data. 
We note, however, that  the present SBT recovers the exact dynamical structure factor (LSWT result) in the large-$S$ limit  {\it for any value of $\alpha$} \cite{Zhang21}. 
Following the procedure described in Ref.~\onlinecite{Ghioldi18}, we use the path-integral formulation.
The auxiliary field $\lambda$ is introduced to enforce the  constraint $ \sum_{\sigma=\uparrow, \downarrow} b_{i \sigma}^{\dag}b_{i \sigma}^{}=1$. To decouple the bilinear forms $\bar X X$
in Eq. (\ref{spinspin}),  we perform a  Hubbard-Stratonovich  transformation,  
\begin{align}
e^{\rm{sgn}(J_{ij})|J_{ij}| \overline  X_{ij}^{} X_{ij}^{} } &= |J_{ij}| \int \frac{ d\overline W_{ij}^{X} \  d W_{ij}^{X}}{2 \pi i} e^{ -|J_{ij}| \overline W_{ij}^{X} W_{ij}^{X} } \nonumber \\
& \times e^{ |J_{ij}| \big( \rm{sgn}(J_{ij})\overline W_{ij}^{X} X_{ij}^{} + W_{ij}^{X} \overline X_{ij}^{} \big) },
\end{align}
where $W^X_{ij}$ with $X=A, B, C,$ and $D$ are the Hubbard-Stratonovich fields. The complex SB or spinon field $b$ can be formally integrated out, and the exact partition function is expressed as a  path integral over the auxiliary fields $W^X$ and $\lambda$,
\begin{equation}\label{Zeff}
 \mathcal{Z}[j] = \int [D\overline W DW][D\lambda] \ e^{-S_{\rm eff}(\overline W, W, \lambda, j,h)}.
\end{equation}
The source $j$ couples the system with a general external magnetic field and it is used to compute correlation functions.
%
The Lagrange multiplier $\lambda$ and the phases of the auxiliary fields $W^X$ are the emergent gauge fields of the SBT~\cite{Ghioldi18}.
The magnetic ordering emerges as a spontaneous  Bose-Einstein condensation of the spinon field.
Since the Hubbard-Stratonovich transformation does not break the U(1) symmetry of ${\mathcal H}$, an
infinitesimal symmetry-breaking field $h$ is necessary to select  a  condensate  associated with a particular choice of the 120$^{\circ}$ ordering (vector chirality and orientation of the ordered moment of a given spin). The effective action can be divided into two contributions: $ S_{\rm{eff}}(\overline W, W, \lambda,j,h) = S_{0}(\overline W, W, \lambda) + S_{\rm bos}(\overline W, W, \lambda,j,h) $, with 
\begin{equation}
 S_{0}({\overline W}, W, \lambda) = \!\! \int_{0}^{\beta} \!\! d\tau \big( \sum\limits_{ij, X} \! J_{ij}^{}  {\overline W}_{ij}^{X \tau} W_{ij}^{X \tau} \!\! - \!\! i 2S \sum_{i} \! \lambda_{i}^{\tau} \big)  ,
\end{equation}
and
\begin{align}
 S_{\rm bos}(\overline W, W, \lambda,j,h) & = - \frac{1}{2} {\rm ln} \int D[\bar b, b] \ e^{-\boldsymbol{b}^\dag \mathcal{M} \ \boldsymbol{b} }  \nonumber \\
 &= \frac{1}{2}  {\rm Tr} \ {\rm ln} \Big[ \mathcal{G}^{-1}(\overline W, W, \lambda, j,h) \Big],
\end{align}
%
where  $\boldsymbol{b}$ is the complex Nambu spinor field, $\beta= 1/ k_B T$, $\mathcal{G}\!=\!\mathcal{M}^{-1}$ is the single-spinon propagator, $\mathcal{M}$ is the bosonic dynamical matrix, and the trace is taken over space, time, and boson indices. The next step is to expand the effective action around its saddle-point (SP) solution (equivalent to the mean-field solution in the canonical formalism), 
\begin{equation}
\label{expand}
    S_{\rm eff} = S_{\rm eff}^{\rm sp} + \sum S_{\alpha_1 \alpha_2}^{(2)} \Delta\phi_{\alpha_1} \Delta\phi_{\alpha_2} + S_{\rm int}
    , 
\end{equation}
where $S_{\rm eff}^{\rm sp}$ is the value of the effective action at the SP solution
$\phi^{sp}_{\alpha}$: $\partial S_{\rm eff} / \partial \phi_{\alpha}|_{\rm sp} = 0$, with  $\phi_{\alpha}  \equiv ( \overline W_{ij}^{X},\ W_{ij}^{X},\  \lambda_{i}^{})$.
The second term is the Gaussian contribution determined by the fluctuation matrix
$S_{\alpha_1 \alpha_2}^{(2)}\!=\! \frac{1}{2} \left( \partial^2 S_{\rm eff} / \partial \phi_{\alpha_1} \partial \phi_{\alpha_2} \right)|_{\rm sp}$
and $\Delta \phi_{\alpha}= \phi_{\alpha} -\phi^{\rm sp}_{\alpha}$.
The third term $S_{\rm int} = \sum_{n=3}^\infty \sum_{\alpha_1 \cdots \alpha_n} S_{\alpha_1 \cdots \alpha_n}^{(n)}  \Delta \phi_{\alpha_1} \cdots \Delta \phi_{\alpha_n}$, with $S_{\alpha_1 \cdots \alpha_n}^{(n)}\!=\! \frac{1}{n!} \left( \partial^n S_{\rm eff} / \partial \phi_{\alpha_1} \cdots \partial \phi_{\alpha_n} \right)|_{\rm sp}$, includes higher-order terms in the fluctuations of the auxiliary fields. \\

At the SP or mean-field level, the %
auxiliary fields $\lambda_i$ and $W_{ij}^X$ are uniform and static. Correspondingly, the mean-field theory describes a noninteracting gas of SBs or spin-$1/2$ spinons with a free-spinon  propagator $\mathcal{G}^{\rm{sp}}$. 
The resulting  condensation of these spinons at $T  \leq T_N$ %
leads to 120$^{\circ}$ magnetic ordering within each triangular layer and antiferromagnetic ordering between adjacent layers~\cite{Arovas88,Auerbach94,Ghioldi2015}. Correspondingly, the free spinon  
propagator acquires a new contribution from the condensate.
Furthermore, as it was shown in recent works~\cite{Ghioldi18,Zhang19}, fluctuations of the auxiliary fields  mediate  spinon-spinon interactions that drastically modify the nature of the low-energy spin excitations  revealed by the dynamical spin susceptibility in Matsubara frequency $i\omega$ and momentum $\boldsymbol{q}$ space:
\begin{equation}
 \chi_{\mu\nu}(\boldsymbol{q},i\omega) = \lim_{h\rightarrow0} \lim_{N_{s}\rightarrow \infty} \frac{\partial^{2} ln \mathcal{Z}^{}[j]}{\partial j_{\boldsymbol{q},i\omega}^{\ \mu} \ \partial j_{-\boldsymbol{q},-i\omega}^{\ \nu}} \bigg|_{j=0} , 
\end{equation}
where $\mu, \nu = x, y, z$ and $N_s$ is the total number of spins. \\
\begin{figure}[t!]
	\centering\includegraphics[width=0.47\textwidth]{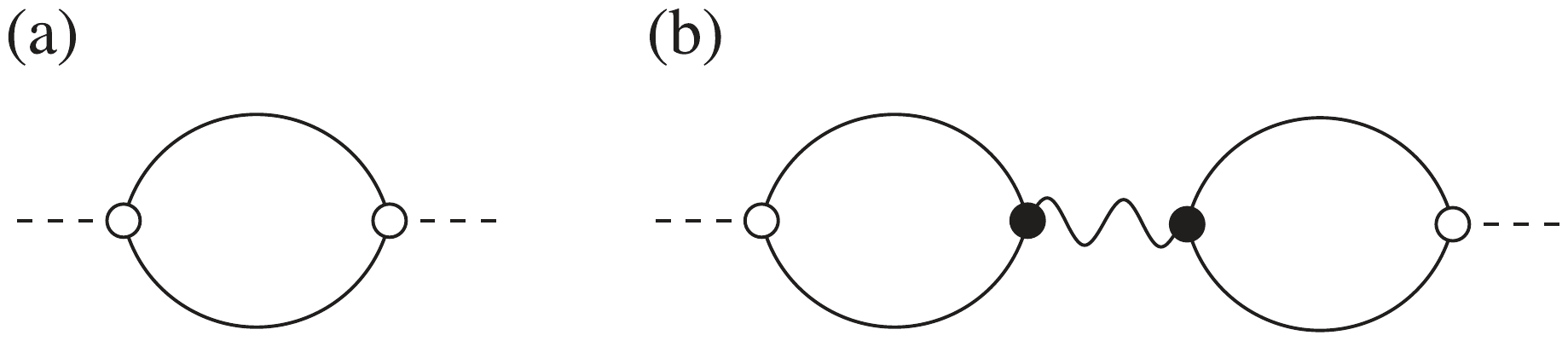}
	\caption{ Diagrammatic representation of different contributions to the dynamical spin susceptibility: 
	(a) saddle-point and  (b) Gaussian fluctuations around the SP solution \cite{Ghioldi18,Zhang21}. The dashed lines represent the external fields. The full lines represent the  single-spinon propagator for the SP solution. The  wavy lines represent the  propagator of the auxiliary fields~\cite{Auerbach94}, whose poles correspond to the  true magnons of the SBT.}
	\label{fig:Feyman}
\end{figure}
By following the procedure described in detail in Ref.~\onlinecite{Ghioldi18}, we compute the $1/N$ correction 
($N$ is the number of bosonic flavors) 
by including the Gaussian fluctuations of the auxiliary fields %
$W^X$ and $\lambda$. At this level, the resulting dynamical spin susceptibility takes the form
\begin{equation}
 \chi_{\mu \nu}(\boldsymbol{q}, i\omega)=\chi_{ \mu\nu}^{\rm sp}(\boldsymbol{q},i\omega)+\chi_{ \mu\nu}^{\rm fl}(\boldsymbol{q},i\omega),
\end{equation}
where 
\begin{equation} \label{chi1sp}
\chi_{\mu\nu}^{\rm{sp}}(\boldsymbol{q},i\omega) = \frac{1}{2} {\rm Tr} \left[ \mathcal{G}^{\rm{sp}}  u^{\mu}(\boldsymbol{q},i\omega) 
 \mathcal{G}^{\rm {sp}} u^{\nu}(-\boldsymbol{q},-i\omega) \right],
\end{equation}
denotes the contribution 
obtained at the saddle-point level, and
\begin{align}
\chi_{ \mu\nu}^{\rm fl}(\boldsymbol{q}, i\omega) =  \sum_{\alpha_1 \alpha_2} &\frac{1}{2} {\rm Tr} \big[ \mathcal{G}^{\rm sp} \ v_{\phi_{\alpha_1}} \ \mathcal{G}^{\rm sp} \ u^{\mu}(\boldsymbol{q}, i\omega) \big] 
\nonumber \\ 
\times D_{\alpha_1 \alpha_2}(\boldsymbol{q},i\omega)&\frac{1}{2} {\rm Tr} \big[ \mathcal{G}^{\rm sp} \ v_{\phi_{\alpha_2}} \ \mathcal{G}^{\rm sp} \ u^{\nu}(-\boldsymbol{q}, -i\omega) \big].  
\label{chiflIIlast}
\end{align}
is the contribution from Gaussian fluctuations around the saddle-point solution.
The propagator of the auxiliary fields $D_{\alpha_1 \alpha_2}(\boldsymbol{q},i\omega)$ (also known as random phase approximation (RPA) propagator) is the inverse of the fluctuation matrix %
$S_{\alpha_1 \alpha_2}^{(2)}$. %
The internal and external vertices, $v_{{\phi}_{\alpha}}\!=\! {\partial \mathcal{G}^{-1}}/{\partial \phi_{\alpha}}$ and  $ u^{\mu}(\boldsymbol{q},i\omega)\! =\! \partial \mathcal{G}^{^{-1}} \! / \partial j_{\boldsymbol{q},i\omega}^{\; \mu} $, couple the spinons to the auxiliary fields $\phi_{\alpha}$  and to the external fields, respectively. %
 The contributions $\chi_{ \mu\nu}^{\rm sp}(\boldsymbol{q},i\omega)$ and $\chi_{ \mu\nu}^{\rm fl}(\boldsymbol{q}, i\omega)$ are represented as Feynman diagrams in  Figs.~\ref{fig:Feyman}(a)  and \ref{fig:Feyman}(b), respectively.

 Historically, the community working on SBT tried to fit experimental results using the mean field susceptibility  $\chi^{\rm{sp}}_{\mu\nu}(\boldsymbol{q},i\omega)$ ~\cite{Auerbach1988,Fak2012,Ghioldi2015,Samajdar2019}. However,  the poles of $\chi^{\rm{sp}}_{\mu\nu}(\boldsymbol{q},i\omega)$
coincide with the poles of the single-spinon propagator $\mathcal{G}^{\rm sp}$ (single-spinon poles). As we demonstrated in Refs.~\onlinecite{Ghioldi18,Zhang19,Zhang21}, the true collective modes (magnons) of the magnetically ordered state arise as two spinon-bound states associated with poles of the propagator of the auxiliary fields. Correspondingly, the magnons of the theory can only be obtained by including contributions from fluctuations around the SP solution.
As it is  discussed in Ref.~\onlinecite{Zhang21}, for each diagram of the $1/N$ expansion of the dynamical spin susceptibility, there is a counter-diagram that cancels the residues of the unphysical single-spinon poles. In particular, the counter-diagram of the SP diagram shown in  Fig.~\ref{fig:Feyman}(a) is the ``fluctuation'' diagram shown in 
Fig.~\ref{fig:Feyman}(b). This seems strange at  first sight because, in absence of a condensate, these diagrams are of different order (the mean field diagram is of order $1/N^0$, while the fluctuation (FL) diagram is of order $1/N$). The key observation is that, \emph{in the presence of a finite condensate fraction,} the second diagram acquires a singular contribution of order $1/N^0$ that cancels the residues of the single-spinon poles of the mean field diagram~\cite{Zhang21}.
The remaining poles arising from the RPA propagator that appears in the second diagram [see Fig.~\ref{fig:Feyman}(b)]  correspond to the true collective modes of the theory. As it was demonstrated in Ref.~\onlinecite{Zhang19}, the energies of the new poles and their spectral weights coincide with the LSWT in the large-$S$ limit.
Among other things, these results explain the failure of previous attempts of recovering the correct large-$S$ limit using a \emph{mean field} SBT~\cite{Chandra1990}.

%
%

\section{Comparison with inelastic neutron scattering experiment}
The total INS cross section at $T=0$ is given by
\begin{equation}\label{Iqw}
    I(\boldsymbol{q}, \omega)= f^2(q) \sum_{\mu} \left( 1 - \frac{q_{\mu}^2}{q^2} \right) S^{\mu \mu}(\boldsymbol{q},\omega),
\end{equation}
where $f(q)$ is the spherical magnetic form factor for Co$^{2+}$ ions, $ \mathcal{S}^{\mu \mu}(\boldsymbol{q}, \omega) = -\frac{1}{\pi} {\rm Im} \left[ \chi_{\mu \mu}(\boldsymbol{q}, \omega)\right] $ is the dynamical spin structure factor, and
 $\chi_{\mu \mu}(\boldsymbol{q}, \omega)$ is the dynamical spin susceptibility computed with the two diagrams shown in Fig.~\ref{fig:Feyman}.\cite{Ghioldi18,Zhang19}\\

\begin{figure*}[t!]
\centering\includegraphics[scale=1.0]{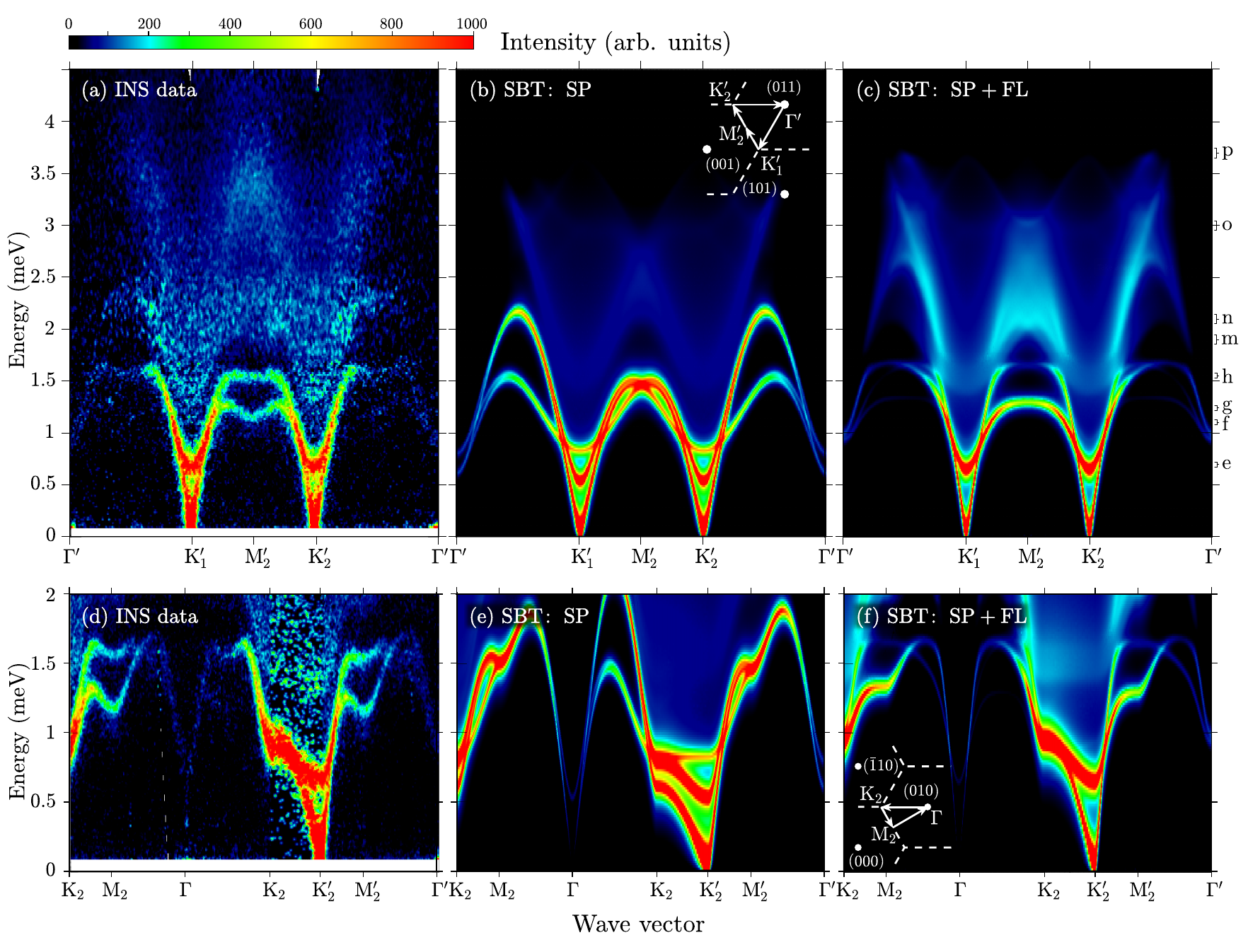}
	\caption{Comparison between the  INS measurements of  Ba$_3$CoSb$_2$O$_9$ reproduced from Ref.~\onlinecite{Macdougal20} (left column) and the zero-temperature $I(\boldsymbol{q},\omega)$ computed with the SBT~\cite{Ghioldi18,Zhang19,Zhang21} (middle and right columns). The middle column shows the mean-field result corresponding to the  diagram depicted in Fig.~\ref{fig:Feyman}(a). The right column includes contributions from both diagrams shown in Fig.~\ref{fig:Feyman}. In all figures included in this work, the calculated $I(\boldsymbol{q},\omega)$ was multiplied by a single overall intensity scale factor to compare with the observed experimental intensity data. 
	The brackets on the right-hand side in (c) indicate the energy cuts shown in Fig. \ref{Energy-cuts}. %
	The wave-vector path in (a)-(c) is $\Gamma' (0,1,1)$ $\rightarrow$ $\rm K_{1}' (1/3,1/3,1)$ $\rightarrow$ $\rm M_2' (0,1/2,1)$ $\rightarrow$ $\rm K_2' (-1/3,2/3,1)$ $\rightarrow$ $\Gamma'$, shown in the inset. In (d)-(f), the path is $\rm K_2 (-1/3,2/3,0)$ $\rightarrow$ $\rm M_{2} (0,1/2,0)$ $\rightarrow$ $\rm \Gamma (0,1,0)$  $\rightarrow$ $\rm K_2$ $\rightarrow$ $\rm K_2' (-1/3,2/3,1)$ $\rightarrow$ $\rm M_{2}' (0,1/2,1)$ $\rightarrow$ $\Gamma' (0,1,1)$ . }
	\label{flo:SchwingerBosons2}
\end{figure*}

\subsection{Single-Magnon Dispersion}

Figures~\ref{flo:SchwingerBosons2}(a) and \ref{flo:SchwingerBosons2}(d) show an overview of 
the measured excitation spectrum of  Ba$_3$CoSb$_2$O$_9$ 
along representative paths in momentum space~\cite{Macdougal20}. In the notation of Ref.~\onlinecite{Macdougal20}, the wave vector labels $\Gamma$, M and K refer to the conventional high-symmetry points in the 2D hexagonal Brillouin zone (BZ), where an unprimed (primed) label indicates $l=0$ ($l=1$) and numbered subscripts refer to symmetry-related distinct points when reduced to the first BZ. 
The scattering intensity is strongest around the magnetic Bragg wave vectors K$^{\prime}_{1,2}$, from which a linearly dispersing in-plane Goldstone mode emerges. The second out-of-plane mode is gapped because of the easy-plane anisotropy. A clear rotonlike minimum appears in the lower-energy mode at the M$^{\prime}_2$ point, 
while the higher-energy mode exhibits a flattened dispersion.

 Figures~\ref{flo:SchwingerBosons2}(b) and \ref{flo:SchwingerBosons2}(e)  include the $T\!=\!0$ INS cross section $I^{\rm{sp}}(\boldsymbol{q},\omega)$ obtained from the SP diagram shown in Fig.~\ref{fig:Feyman}(a). As anticipated in the previous section, the poles of $\chi^{\rm{sp}}_{\mu\nu}(\boldsymbol{q},i\omega)$ coincide with the poles of the single-spinon propagator $\mathcal{G}^{\rm sp}$  because they arise from replacing one of the two propagators in the Feynman diagram with the contribution from the condensate $\mathcal{G}^{\rm sp}_c$. In addition to the single-spinon poles,  $I^{\rm{sp}}(\boldsymbol{q},\omega)$ exhibits a W-shaped continuum scattering [see Fig. \ref{flo:SchwingerBosons2}(b)]  arising from the two-spinon continuum, which extends up to twice the single-spinon bandwidth: $2 W_{\rm spinon} \simeq 4.32$ meV. 

Figures~\ref{flo:SchwingerBosons2}(c) and \ref{flo:SchwingerBosons2}(f) show the $T\!=\!0$ INS cross section $I(\boldsymbol{q},\omega)$ obtained from the sum of the $\rm SP$ and $\rm FL$
diagrams  included in Fig.~\ref{fig:Feyman}. The addition of the counterdiagram depicted in Fig.~\ref{fig:Feyman}(b) changes the result at a qualitative level. As anticipated, it cancels out the residues of the single-spinon poles of $\chi^{\rm sp}_{\mu \mu}(\boldsymbol{q}, \omega)$, implying that the only poles of the resulting  $\chi_{\mu \mu}(\boldsymbol{q}, \omega)$ are the poles of the RPA propagator [wavy line in Fig.~\ref{fig:Feyman}(b)]. These poles correspond to the single-magnon excitations, which are the true collective modes of the system. In addition, the W-shaped two-spinon continuum, shown in Fig. \ref{flo:SchwingerBosons2}(c), becomes more pronounced, exhibiting larger intensity and a stronger modulation as a function of energy and momentum.

\begin{figure}[t!]
	\centering\includegraphics[scale=0.95]{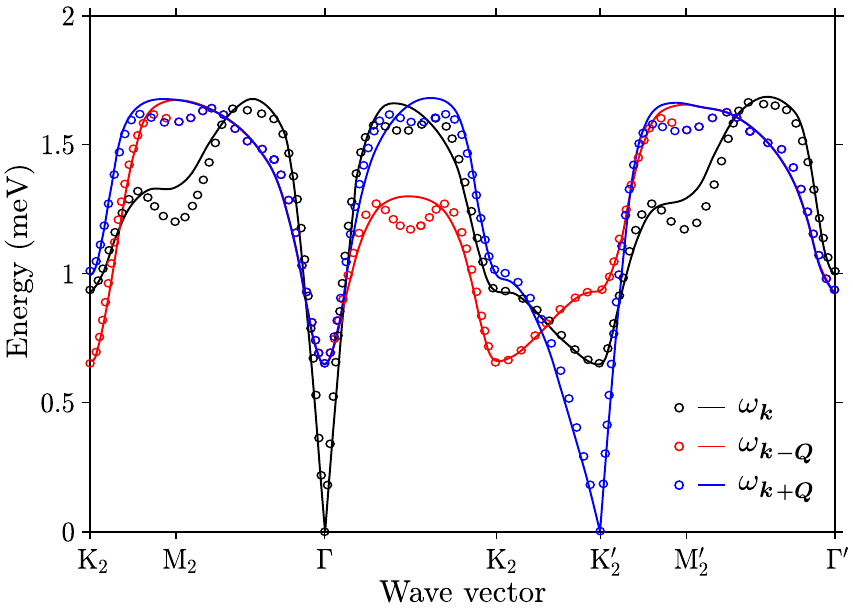}
	\caption{Comparison between the two-spinon bound state  dispersion computed with SBT (full lines) and the best parametrization (open circles) of the experimentally observed magnon dispersions from Ref.~\onlinecite{Macdougal20}. }
	\label{fig:mag}
\end{figure}

The failure of  NLSWT has motivated an empirical parametrization of the single-magnon dispersion with more than 10 fitting parameters~\cite{Macdougal20}. Figure \ref{fig:mag} includes a comparison between this experimentally fitted single-magnon dispersion and the  single-magnon dispersion  extracted from the poles of the RPA propagator.
The SBT reproduces the measured magnon dispersion to a very good approximation. Remarkably, the only tuning parameter is $\alpha\!=\!0.436$, which turns out to be very close to the value $\alpha\!=\!0.5$ adopted in previous works~\cite{Trumper1997,Manuel1998,Manuel99,Ghioldi18,Zhang19}. The comparison reveals that the overall single-magnon dispersion
is very well reproduced by the 
SBT, which predicts a magnon velocity $ c_m \approx 1.2 J$. The only noticeable discrepancies are the small rotonlike anomalies near the $\rm M^{}$  and $\rm K^{}/2$
points.  Returning to Fig.~\ref{flo:SchwingerBosons2}, the overall spectral weight modulation of the sharp magnons is also well reproduced  over the whole Brillouin zone, except for the points that exhibit the rotonlike anomaly.
This level of agreement is remarkable if we consider that NLSWT predicts a single-magnon bandwidth of  2.4 meV, which is more than 40$\%$ higher than the experimental value~\cite{Ma16}.
The combination of both results suggest that a free-spinon gas is a better starting point to describe the magnons of Ba$_3$CoSb$_2$O$_9$, which arise as two-spinon bound states in the SBT. 

\begin{figure}[t!]
	\centering\includegraphics[scale=0.97]{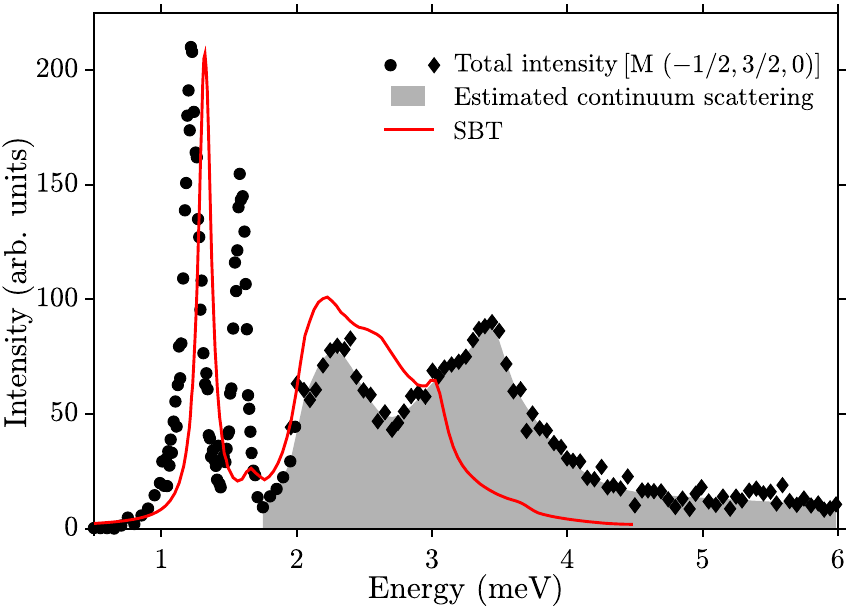}
	\caption{Comparison between  INS intensity averaged %
	over $l$ (including several zones as in Ref.~\onlinecite{Macdougal20})
	and the corresponding average of the intensity $I(\boldsymbol{q}, \omega)$  obtained from the two diagrams shown in Fig.~\ref{fig:Feyman} (solid line).  A Lorentzian broadening, which accounts for  the estimated experimental energy resolution, has been introduced in the calculated $I(\boldsymbol{q}, \omega)$ via the replacement $\omega \to \omega+ i \eta$.    }
	\label{flo:SchwingerBosons3}
\end{figure}

The lack of the rotonlike anomalies and the corresponding renormalization of the single-magnon spectral weight are expected shortcomings of the current level of approximation, if we consider that the diagrams shown in Fig.~\ref{fig:Feyman} correspond to the lowest-order approximation  required to obtain the true collective modes of the theory. In other words, these diagrams
do not include self-energy corrections of the single-spinon and the auxiliary field propagators. It is well known that rotonlike anomalies arise in NLSWT only after including self-energy $1/S$ corrections to the bare single-magnon propagator~\cite{Starykh06,Chernyshev06,Zhitomirsky13,Mourigal2013}. In the case of the SBT, self-energy corrections to the single-spinon propagator also renormalize the single-magnon dispersion because magnons are two-spinon bound states. This renormalization is  expected to shift the position of the magnon peaks relative to the onset of the two-spinon continuum. As shown in Fig.~\ref{flo:SchwingerBosons3}, the overlap between the higher-energy magnon at the $\rm M'$ point and the two-spinon continuum leads to a strong reduction of the spectral weight, which is not observed in the experiment, where the separation between the magnon peak and the continuum is roughly 0.3 meV [see Fig.~\ref{flo:SchwingerBosons2}(a)]. Based on these observations, we conjecture that
the rotonlike anomalies will arise from self-energy corrections of the single-spinon and/or single-magnon propagators.\\ 

\begin{figure*}[t]
	\centering\includegraphics[width=0.95\textwidth]{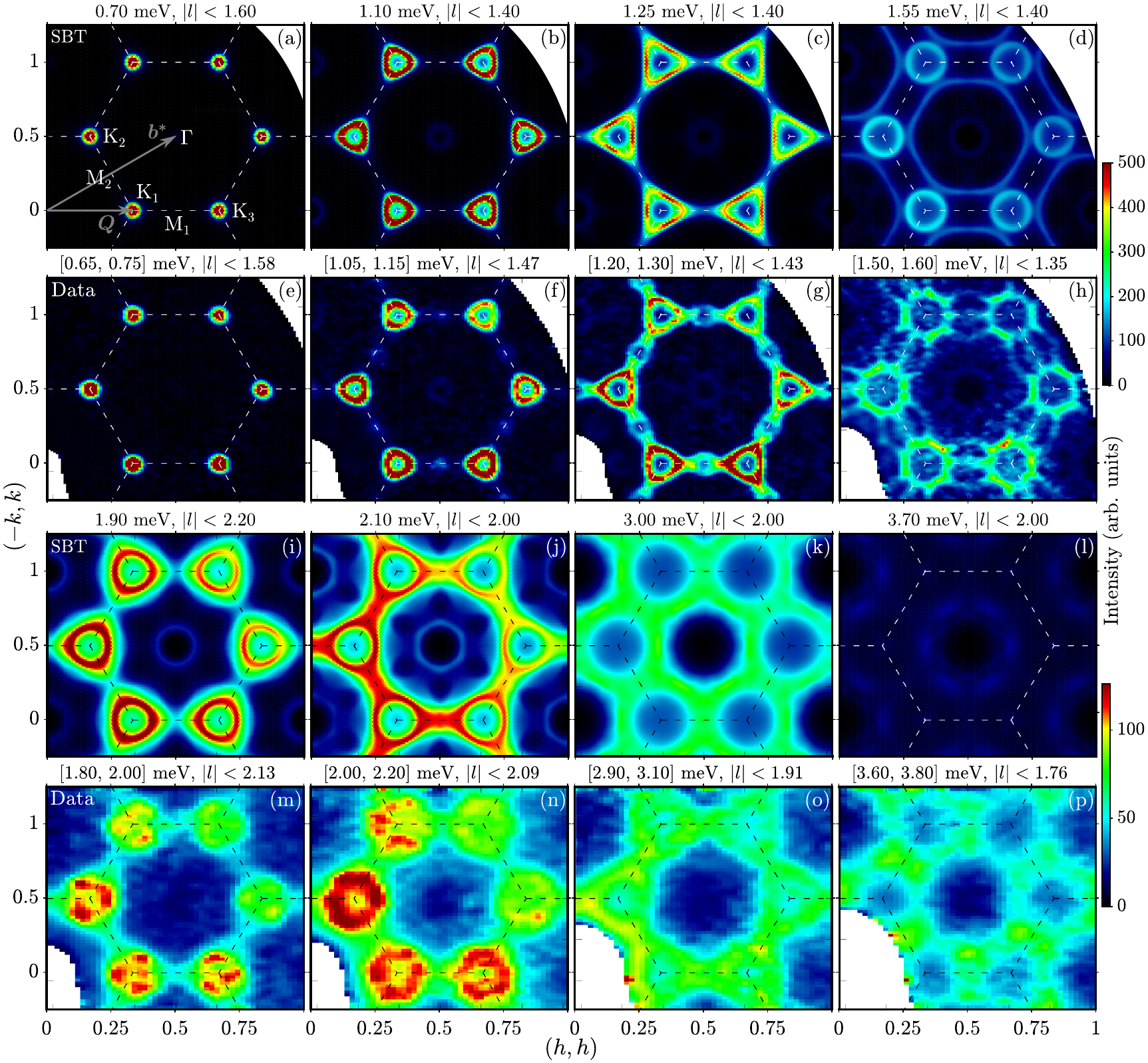}
	\caption{Intensity maps of $I({\bf q}, \omega)$ [Eq. (\ref{Iqw})] as a function of momentum in the $hk$ plane at a series of constant energies. The results have been integrated over an energy range that is indicated at the top of each panel to facilitate the comparison  with INS data reproduced from Ref.~\onlinecite{Macdougal20}.}
	\label{Energy-cuts}
\end{figure*}

\subsection{Continuum Scattering} 

Another important consequence of the composite nature of the single-magnon excitations is the emergence of a highly structured two-spinon continuum,   which extends up to twice the single-spinon bandwidth: $2 W_{\rm spinon} \simeq 4.32$ meV. 
As it is clear from Figs.~\ref{flo:SchwingerBosons2}(b) and \ref{flo:SchwingerBosons2}(c), the single-spinon bandwidth $W_{\rm spinon} \simeq 2.16$ meV is significantly larger than the single-magnon bandwidth $W_{\rm magnon} \simeq 1.7$ meV.
While this difference could explain the origin of the  wide energy window of continuum scattering revealed by the INS experiment, we will see below that the SBT theory  is still missing spectral weight in the high-energy region at the current level of approximation.

The SBT reproduces the strong intensity modulations and the dispersion across the Brillouin zone of the low-energy part of the continuum  [see Figs.~\ref{flo:SchwingerBosons2}(a) and \ref{flo:SchwingerBosons2}(c)]. %
For instance, Fig.~\ref{flo:SchwingerBosons3} shows the average over $l$ of the experimental and theoretical neutron scattering cross section at $(1/2,1/2,l)$. The theoretical ratio  between the %
INS intensity
of the continuum and the  magnon peaks is approximately equal to   $2.74$, which is in remarkably good agreement with the value of 2.8 that is  obtained by integrating the experimental curve in  Fig.~\ref{flo:SchwingerBosons3}. Importantly, this ratio is more than four times  higher than the value of $\simeq 0.66$ obtained from NLSWT~\cite{Kamiya18}, which clearly underestimates the relative weight of the continuum scattering.
Moreover, the measured continuum scattering   extends up to at least $E \simeq$ 6~meV, which is roughly equal to four times 
the single-magnon bandwidth~\cite{Ito2017,Macdougal20}. 
The two diagrams included in our SBT  only account for the first two low-energy stages.
We attribute this discrepancy to the lack of four-spinon contributions arising from self-energy corrections to the single-spinon propagator not included in Fig.~\ref{fig:Feyman}. 

To facilitate the comparison with the the INS data~\cite{Macdougal20}, Fig.~\ref{Energy-cuts} shows intensity maps of the measured and calculated $I({\boldsymbol q}, \omega)$ at different energies.
The comparison at energies below the top of the magnon band, shown in Figs. \ref{Energy-cuts}(a)-\ref{Energy-cuts}(h), confirms the above-discussed overall agreement between experiment and theory, except for the missing rotonlike anomalies in the theoretical calculation, that explain the differences between the Figs. \ref{Energy-cuts}(c) and Figs. \ref{Energy-cuts}(g) near the M points. 

Figures \ref{Energy-cuts}(i)-\ref{Energy-cuts}(p) show the intensity maps arising from the continuum scattering just above the top of the one-magnon dispersion. As observed in the experiment, the continuum intensity obtained from the SBT is  centered at the K points with a clear threefold symmetric pattern. The ring patterns around K, that become apparent at slightly higher energies and transform into triangular contours with corner touchings at the M points, are also reproduced by the SBT. However, the ring patterns of the theoretical calculation are rotated by an angle $\pi/3$ relative to the ring patterns of the experimental data [see Figs. \ref{Energy-cuts}(i) and \ref{Energy-cuts}(m)]. Once again, we attribute this difference to the absence of  the  rotonlike anomaly in the theoretical calculation. 
As shown in Fig.~\ref{flo:SchwingerBosons3}, %
the neutron scattering intensity of the measured upper magnon peak near the M points is significantly larger than the calculated %
 intensity. It is then  expected that self-energy corrections to the single-spinon propagator, which account for the rotonlike anomaly, should  transfer spectral weight from the low-energy continuum (around $E \simeq 2$ meV) to the upper magnon peak. The excess of continuum spectral weight near the M points at the current level of approximation explains the $\pi/3$ rotation of the ring patterns and the ``bridges'' that connect adjacent rings in Fig. \ref{Energy-cuts}(j), which do not have a counterpart in the experimental data shown in Figs. \ref{Energy-cuts}(n). Finally, as it is clear from the comparison between Fig. \ref{Energy-cuts}(l) and \ref{Energy-cuts}(p), the relatively large spectral weight 
of the measured continuum scattering in the high-energy interval ranging from 3.6 to 3.8 meV is not reproduced by the SBT at the current level of approximation (see  Fig.~\ref{flo:SchwingerBosons3}).

\section{Discussion}  

Our detailed comparison between the SBT and the INS cross section of Ba$_3$CoSb$_2$O$_9$ reveals, for the first time, that a low-order expansion in the control parameter ($1/N$) provides
an adequate framework to describe the magnetic excitations of quasi-$2\rm D$ TLHA. 
 In contrast,  semiclassical treatments overestimate the single-magnon bandwidth by approximately 40\% and they cannot account for the large intensity and modulation of the observed continuum scattering~\cite{Ma16,Ito2017}.
Thus, we attribute the failure of the large-$S$ expansion to the proximity of Ba$_3$CoSb$_2$O$_9$ to a QCP that signals the onset of a QSL. Since the elementary excitations of the QSL phase are quasifree single spinons, a free-spinon gas becomes  a better starting point than a free-magnon gas near the QCP. Magnons  are then recovered on the magnetically ordered side  as two-spinon bound states (poles of the RPA propagator) induced by  fluctuations of the emergent gauge fields.
Within the SBT, the gapped $\mathbb{Z}_2$ QSL state proposed by Sachdev~\cite{Sachdev92} is the only liquid which  can be continuously connected with a $120^{\circ}$ N\'eel ordered state~\cite{Wang06}, as it does not break any symmetries and has its lowest energy modes at the $\rm K$ points. The resulting quantum critical point is expected to have a dynamically generated $O(4)$ symmetry~\cite{Azaria90,Chubukov94}.

Alternative parton theories with fermionic matter fields lead to a different spin liquid state on the other side of the QCP, such as a gapless U(1) spin liquid~\cite{Dupuis19,Hu19}. However, while  existing attempts  to reproduce the unusual excitation spectrum of the ordered phase using fermionic  partons seem to account for the rotonlike anomaly, the results have not been  compared against the available experimental data~\cite{Zhang2020,Ferrari19}.

Extended continua has also been observed in ladder \cite{Lake2009} and spatially anisotropic triangular \cite{Kohno2007} systems 
 -experimentally realized in CaCu$_2$O$_3$ and Cs$_2$CuCl$_4$ compounds
\cite{Lake2009,Kohno2007}, respectively. 
These continua have been attributed to 1D spinons which are confined by the interchain interactions. This is in sharp contrast to the 2D character of the spinons invoked in this work, which are the building blocks of the SBT and interact via emergent gauge fields consisting of the Lagrange multiplier and phases of the bond fields $W_{ij}^{X}$.

Our results have implications for other quantum magnets that are described by a similar model. For instance, the delafossite triangular lattice materials, such as CsYbSe$_2$~\cite{xie2021field} and NaYbSe$_2$, 
could lie even closer to the quantum melting point, while the triangular layers of Ba$_2$CoTeO$_6$~\cite{Kojima2022} exhibit an INS spectrum that is remarkably similar to the one of Ba$_3$CoSb$_2$O$_9$. 
A very recent tensor network study of the triangular \textit{XXZ} model \cite{chi22} reinforces the validity of this model to quantitatively describe the magnetic excitations of Ba$_3$CoSb$_2$O$_9$.  Recently, we also became aware of Ref.~\onlinecite{syromyatnikov21}, which attempts to solve the same problem using a different approach.  \\

\section{Acknowledgments}

We thank Radu Coldea for a critical reading of our manuscript and for providing detailed explanations of the data presented in Ref.~\onlinecite{Macdougal20}. We also  acknowledge useful discussions with %
 D. A. Tennant, A. Scheie, C. J. Gazza, O. Starykh, Alexander Chernyshev, and M. Mourigal.
The work by C.D.B.  was supported by the U.S. Department of Energy, Office of Science, Basic Energy Sciences, Materials Sciences and Engineering Division under Award No. DE-SC-0018660.
Y.K.~acknowledges the support by the NSFC (Grants No.~12074246 and No.~U2032213) and MOST (Grants No.~2016YFA0300500 and No.~2016YFA0300501) research programs.  E.A.G., L.O.M and A.E.T. were supported by CONICET under Grant PIP No. 3220.
\bibliography{Ref2}

\end{document}